\documentclass[]{aa}
\usepackage{txfonts} 
\usepackage{graphicx}
\usepackage{natbib}

\newcommand{\Teff}{$T_\mathrm{eff}$}
\newcommand{\logg}{$\log~g$}

\begin{document}

\title{Cool DZ white dwarfs in the SDSS}

\author{D. Koester \inst{1}
   \and    J. Girven  \inst{2}
   \and    B.T. G\"ansicke \inst{2}
   \and    P. Dufour   \inst{3}  }

\offprints{D. Koester}

\institute{
          {Institut f\"ur Theoretische Physik und Astrophysik, 
             University of Kiel, {D-24098 Kiel}, Germany\\
          \email{koester@astrophysik.uni-kiel.de}}
   \and  {Department of Physics, University of Warwick, Coventry, UK}
   \and  {D\'epartement de Physique, Universit\'e de Montr\'eal, 
          Montr\'eal, QC H3C 3J7, Canada}
   }

\date{}

\abstract{}{We report the identification of 26 cool DZ white dwarfs
  that lie across and below the main sequence in the Sloan Digital Sky
  Survey (SDSS) $(u-g)$ vs. $(g-r)$\ two-color diagram; 21 of these
  stars are new discoveries. }{The sample was identified by visual
  inspection of all spectra of objects that fall below the main
  sequence in the two-color diagram, as well as by an automated search
  for characteristic spectral features over a large area in color
  space that included the main sequence. The spectra and photometry
  provided by the SDSS project are interpreted with model atmospheres,
  including all relevant metals. Effective temperatures and element
  abundances are determined, while the surface gravity has to be
  assumed and was fixed at the canonical value of \logg\ = 8.}{These
  stars represent the extension of the well-known DZ sequence towards
  cooler temperatures and fill the gap around \Teff\ = 6500~K present
  in a previous study. The metal abundances are similar to those in
  the hotter DZ, but the lowest abundances are missing, probably
  because of our selection procedures. The interpretation is
  complicated in terms of the accretion/diffusion scenario, because we
  do not know if accretion is still occurring or has ended long
  ago. Independent of that uncertainty, the masses of the metals
  currently present in the convection zones -- and thus an absolute
  lower limit of the total accreted masses -- of these stars are
  similar to the largest asteroids in our solar system.}{}


\maketitle

\section{Introduction}

Cool white dwarfs with traces of metals other than carbon are
designated with the letter ``Z'' in the current classification
system. If Balmer lines are visible, they are called DAZ or DZA,
depending on whether hydrogen lines or some metal lines are the
strongest. Helium-dominated DBs with metals are DBZ, and finally, at
lower temperature where no hydrogen or helium lines remain visible,
there are the DZs. Helium-rich DBZs and DZs have been known for a long
time; in fact, one of the first three classical white dwarfs, van
Maanen 2, belongs to this class \citep{Kuiper41}. DAZs were
discovered much later \citep{Koester.Provencal.ea97,
  Holberg.Barstow.ea97}, but this is purely observational bias, since
the metal line strengths are much weaker in hydrogen atmospheres with
lower transparency.

The almost mono-elemental composition of the outer layers is caused by
gravitational settling, with time scales much shorter than the white
dwarf evolutionary time scales. The metals can therefore not be
primordial. For decades the standard explanation has been accretion
from interstellar matter (ISM), with subsequent diffusion downward out
of the atmosphere. This scenario has been discussed in detail and
compared with observations in a series of three fundamental papers by
\cite{Dupuis.Fontaine.ea92,Dupuis.Fontaine.ea93,Dupuis.Fontaine.ea93*b}.
A special case is the DQ spectral class, with pollution by the element
carbon. The carbon can be dredged-up from below the outer helium layer
by a deepening convection zone and does not necessarily need an
exterior source \citep{Koester.Weidemann.ea82,
  Pelletier.Fontaine.ea86}.

The ISM accretion hypothesis for the DZ has a number of problems, the
most disconcerting ones being the lack of dense interstellar clouds in
the solar neighborhood, and in many cases a large deficit of hydrogen
in the accreted matter.  These arguments have been summarized by
\cite{Farihi.Barstow.ea10}, who in the same paper argue for the
accretion of circumstellar dust as the source of the accreted
matter. This dust is thought to originate from the the tidal
disruption of some rocky material, left over from a former planetary
system \citep{Debes.Sigurdsson02, Jura03}. This idea has gained more
and more acceptance in the last decade, mainly through the discovery
of infrared excesses due to circumstellar dust in a significant
fraction of DAZs and DBZs \citep{Jura06, von-Hippel.Kuchner.ea07,
  Farihi.Jura.ea09, Farihi.Jura.ea10, Farihi.Barstow.ea10}. In
particular the gaseous disks with \ion{Ca}{ii} emission lines
\citep{Gansicke.Marsh.ea06, Gansicke.Marsh.ea07,
  Gansicke.Koester.ea08} are the strongest confirmation for the disk
geometry and its location inside the Roche lobe of the white dwarfs, a
key prediction of the tidal disruption hypothesis.

A sample of 147 cool He-rich metal polluted white dwarfs (spectral
class DZ) identified by SDSS has been analyzed by
\cite{Dufour.Bergeron.ea07}. Of these, only two have \Teff\ below
6600~K (at 6090 and 4660~K). Given that already about half a dozen
cool DZ known within 20\,pc of the Sun are brighter than this SDSS
sample (including van~Maanen~2 at only 4.3\,pc,
\citealt{sionetal09-1}) this suggests that a large number of cooler DZ
remain to be found.

\section{Color selection}

We have been investigating the spectroscopic data base for white
dwarfs with unusual properties \citep{gaensickeetal10-1}, and
identified SDSS0916+2540 (which we adopt as notation for SDSS objects
throughout this paper) as a cool, very metal-rich DZ white dwarf. In
the $(u-g)$ vs. $(g-r)$ color-color diagram, this star lies below the
main sequence, a region that has hitherto not been systematically
explored for its stellar content. SDSS intensively targeted this
region as part of its quasar program
\citep{richardsetal02-1}. Inspecting all spectroscopic objects within
SDSS Data Release\,7 \citep{abazajianetal09-1} that have colors within
the dashed region shown in Fig.\,1 confirms that the vast majority are
$z\sim3$ quasars, however, we also identified 16 of these objects as
cool DZ white dwarfs characterized by a strong complex of MgI
absorption lines near 5170~\AA\ (one of these turned out to be a
carbon-rich DQ, see below). This finding, coupled with the fact that
all except one of the SDSS DZ stars analyzed by
\citet{Dufour.Bergeron.ea07} are located above the main sequence in
$(u-g)$ vs. $(g-r)$ (Fig.\,1) strongly suggested that part of the DZ
population will fall right into the color space occupied by the
main sequence. In the view of the large number of DR7 spectra of
$\sim$K-A type stars we developed a search routine in the region
bounded by the solid lines in Fig.\,1, which checks for the
characteristic spectral features, foremost the \ion{Mg}{i} lines near
5170~\AA. Secondary criteria were the presence of the Na~D lines and
the absence of H$\alpha$. A heavy contamination are still K stars with
molecular bands of MgH, but since the slope of the spectrum is quite
different, they can be identified by visual inspection. This procedure
led to the discovery of nine additional DZ white dwarfs, two of which were
already in the Dufour sample. Two additional objects, SDSS0143+0113
and SDSS2340+0817, were identified independently by Dufour in a search
for DZs. The complete list of cool DZ (plus one DQ) white dwarfs
identified and their $ugriz$ photometry is given in
Table~\ref{objects}.

\begin{figure}
  \includegraphics[width=0.45\textwidth]{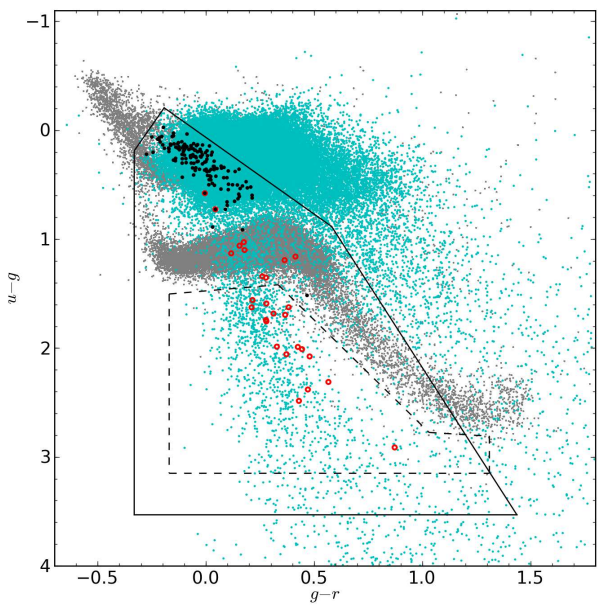} 

\caption{ SDSS Two-color diagram for objects with spectra identified
  as normal stars (grey S-shaped region in the middle of the figure),
  QSO (cyan dots), DZ white dwarfs analyzed by
  \cite{Dufour.Bergeron.ea07} (solid black dots) and cool ``extreme''
  DZ discovered by our search (red open circles).}
\end{figure}

\begin{figure}
  \includegraphics[width=0.45\textwidth]{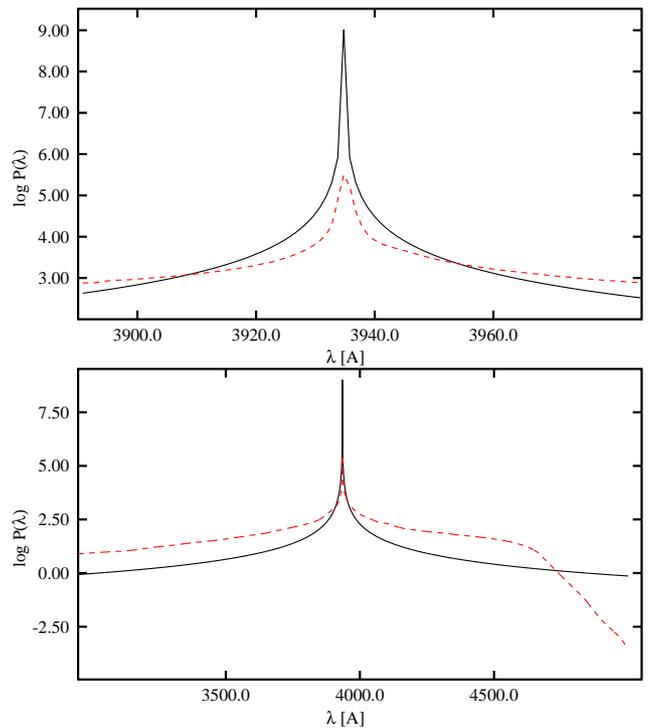} 
\caption{Quasi-static line profile for the \ion{Ca}{ii}\,K resonance
  line (dashed, red), compared with the impact approximation
  (continuous, black). Parameters are 6000~K for the temperature and a
  neutral He perturber density of $10^{19}$cm$^{-3}$. The central
  region is enlarged in the top panel. The quasi-static calculation is
  only useful for the wing; approximately within the central
  region between the crossing points of the two profiles the impact
  calculation is the superior approximation.
\label{profile}}
\end{figure}

\begin{table*}
\caption{DZ white dwarfs in the sample \label{objects}}
\centering          
\begin{tabular}{c c c c c c c}     
\hline\hline    
\noalign{\smallskip}   
SDSS & plate-MJD-fib & $u$ & $g$ &  $r$ & $i$ & $z$ \\
\hline                    
\noalign{\smallskip}
014300.52+011356.8& 53265-1907-405&20.410 (0.060)&19.380 (0.010)&19.200 (0.010)&19.260 (0.020)&19.390 (0.060)\\
015748.14+003315.0& 52199-0700-627&21.261 (0.096)&19.566 (0.020)&19.120 (0.017)&19.237 (0.027)&19.353 (0.053)\\
020534.13+215559.7& 53349-2066-223&21.076 (0.097)&19.916 (0.019)&19.502 (0.017)&19.446 (0.021)&19.430 (0.057)\\
091621.36+254028.4& 53415-2087-166&21.258 (0.102)&18.346 (0.013)&17.473 (0.016)&17.384 (0.016)&17.528 (0.019)\\
092523.10+313019.0& 53379-1938-608&21.039 (0.105)&18.983 (0.019)&18.611 (0.015)&18.632 (0.015)&18.748 (0.045)\\
093719.14+522802.2& 53764-2404-197&20.780 (0.067)&19.439 (0.029)&19.178 (0.019)&19.301 (0.022)&19.370 (0.055)\\
095645.14+591240.6& 51915-0453-621&18.971 (0.024)&18.394 (0.018)&18.399 (0.021)&18.554 (0.020)&18.743 (0.040)\\
103809.19$-$003622.4&51913-0274-265&17.710 (0.020)&16.984 (0.021)&16.940 (0.014)&17.087 (0.013)&17.308 (0.017)\\
104046.48+240759.5& 53770-2352-004&21.717 (0.147)&19.337 (0.025)&18.866 (0.018)&18.870 (0.023)&19.102 (0.063)\\
104319.84+351641.6& 53431-2025-328&20.642 (0.077)&18.959 (0.020)&18.647 (0.020)&18.611 (0.017)&18.665 (0.038)\\
110304.15+414434.9& 53046-1437-410&21.718 (0.186)&19.731 (0.027)&19.403 (0.020)&19.385 (0.021)&19.425 (0.057)\\
114441.92+121829.2& 53138-1608-515&20.723 (0.085)&18.412 (0.018)&17.846 (0.019)&17.755 (0.019)&17.758 (0.022)\\
115224.51+160546.7& 53415-1762-522&21.728 (0.131)&20.169 (0.025)&19.954 (0.021)&20.012 (0.031)&20.103 (0.092)\\
123415.21+520808.1& 52379-0885-269&19.545 (0.042)&18.415 (0.024)&18.299 (0.018)&18.424 (0.015)&18.681 (0.032)\\
133059.26+302953.2& 53467-2110-148&18.300 (0.022)&16.310 (0.016)&15.885 (0.012)&15.898 (0.025)&16.095 (0.024)\\
133624.26+354751.2& 53858-2101-254&19.481 (0.027)&17.854 (0.015)&17.643 (0.013)&17.717 (0.027)&17.827 (0.018)\\
135632.63+241606.0& 53792-2119-216&20.351 (0.057)&18.998 (0.026)&18.721 (0.016)&18.741 (0.019)&18.814 (0.044)\\
140410.72+362056.8& 54590-2931-302&20.010 (0.041)&18.818 (0.029)&18.454 (0.018)&18.520 (0.021)&18.692 (0.033)\\
140557.10+154940.5& 54272-2744-586&20.000 (0.043)&18.940 (0.020)&18.785 (0.019)&18.901 (0.018)&19.104 (0.042)\\
142120.11+184351.6& 54533-2773-088&22.096 (0.168)&20.355 (0.031)&20.077 (0.024)&20.194 (0.037)&20.329 (0.110)\\
143007.15$-$015129.5&52409-0919-358&21.485 (0.141)&19.475 (0.021)&19.031 (0.021)&18.993 (0.019)&19.090 (0.061)\\
152449.58+404938.1& 54626-2936-537&21.760 (0.120)&20.135 (0.028)&19.752 (0.021)&19.765 (0.027)&19.808 (0.078)\\
153505.75+124744.2& 54240-2754-570&18.054 (0.024)&15.977 (0.020)&15.497 (0.018)&15.452 (0.022)&15.504 (0.017)\\
154625.33+300946.2& 53142-1390-362&21.398 (0.085)&19.807 (0.020)&19.528 (0.018)&19.647 (0.023)&19.794 (0.063)\\
155429.01+173545.9& 53875-2170-154&18.698 (0.024)&17.599 (0.015)&17.420 (0.011)&17.468 (0.011)&17.650 (0.023)\\
161603.00+330301.2& 53239-1684-340&21.006 (0.095)&19.252 (0.018)&18.973 (0.020)&19.069 (0.019)&19.154 (0.045)\\
234048.74+081753.3& 54326-2628-108&22.733 (0.273)&20.248 (0.020)&19.819 (0.020)&19.871 (0.027)&19.953 (0.075)\\
\hline                  
\end{tabular}
\tablefoot{Spectral identifiers, and SDSS photometry $ugriz$
  with errors in parentheses. SDSS0142+0113, SDSS0937+5228,
  SDSS1404+3620, and SDSS1524+4049 have a second spectrum of slightly
  lower quality in the SDSS archive. SDSS0937+5228 (WD0933+526),
  SDSS0956+5912 (WD 0953+594), SDSS1038$-$0036 (WD 1035$-$003),
  SDSS1330+3029 (WD1328+307), and SDSS1535+1247 (WD1532+129, G137-24)
  are known DZ (see text for details), the others are new
  identifications of this study.}
\end{table*}

\section{Input physics and models}
\subsection{Line identification and broadening}
All spectral features found in the spectra of the new cool DZ white
dwarfs can be identified with lines from Ca, Mg, Na, Fe, Ti, and Cr,
broadened predominantly through van der Waals broadening by neutral
helium.  The broadest lines show strongly asymmetric profiles, which
were in the case of \ion{Mg}{i} 5169/5174/5185 originally identified
as due to quasi-static broadening by \cite{Wehrse.Liebert80} in their
study of SDSS1330+3029 (= G165-7). The same conclusion was reached by
\cite{Kawka.Vennes.ea04} for SDSS1535+1247; both of these stars are in
our present sample.  The width of these lines, as well as that of the
\ion{Ca}{ii} resonance lines is far beyond the range of validity of
the impact approximation, which is in these cases approximately
8-10~\AA\ from the line centers \citep[see p. 312 in][]{Unsold68}.  We
have used the simple and elegant method of \cite{Walkup.Stewart.ea84},
who present numerical calculations for the transition range between
impact and quasi-static regime, which can in both limits be easily
extended with the asymptotic formulae. These profiles are reasonable
approximations for the MgI triplet.

For the much wider \ion{Ca}{ii} resonance lines this approximation
fails. The reason is very likely that for such strong interactions
needed to produce a 600~\AA\ wide wing the approximation with a
simple van der Waals $r^{-6}$ law is not valid. We have used the
quasi-static limit of the semiclassical quasi-molecular broadening
theory as described in \cite{Allard.Kielkopf82}. This formulation
(e.g. their eq.~59 in the cited paper) easily allows the incorporation
of a Boltzmann factor to account for the variation of the perturbation
probability with distance of the perturber in thermal equilibrium. It
would also allow us to take a variation in the dipole moments into
account, which, however, are apparently not available in the published
literature.

Adiabatic potential energy curves for the ground state of the Ca$^+$He
quasi-molecule and the two exited states correlated with the resonance
term of the \ion{Ca}{ii} ion -- (4p)$^2\Sigma$ and (4p)$^2\Pi$ -- were
calculated by \cite{Czuchaj.Rebentrost.ea96} and numerical data were
presented in a table. Approximate calculations for the spin-orbit
interaction show a mixing of the two upper levels of the doublet and a
complicated structure of the energy curves (their Fig.~4). Since no
numerical data are given for these calculations, we use the potential
curves without this interaction.

These profiles (Fig.~\ref{profile}) were used in the calculation of
models and provide a fairly good fit for the red wing; the blue wing
is mostly outside the SDSS wavelength range. Nevertheless, the current
analysis is only a first attempt with a number of shortcomings:
\begin{itemize}
\item A more accurate description of the line profile needs to take
  into account the spin-orbit interaction and the variation of the
  dipole moments.
\item The potential curves are calculated only to the closest distance
  of $\approx $ 1.6~\AA, since the authors were interested in possible
  minima and bound states of the molecule. The distant wings of the
  profile depend somewhat on how the potential is extrapolated to
  shorter distances.
\item Since the \ion{Ca}{ii} ion has a net charge, the interaction
  with the He atom has a polarization term with a $r^{-4}$ dependence
  on separation, which dominates at large distances.  This is
  different from the case of a neutral radiator, where the potential
  is the van der Waals $r^{-6}$ type. However, since the polarization
  term is the same for upper and lower state, it cancels out in the
  difference, which is the important quantity for line broadening. In
  principle the $r^{-6}$ term should remain, but might be distorted by
  numerical noise. To avoid that, we have replaced the potential
  difference at large distances by an accurate van der Waals term,
  which allows us to recover a reasonable profile also near the line
  center.
\item The quasi-static calculation is valid only in the wings. Within
  approximately 20~\AA\ of the line center the impact approximation
  would be a better description.  A complete calculation of the
  unified profile similar to the calculations for the Na and K
  resonance lines by \cite{Allard.Allard.ea03} would be
  desirable. Nevertheless, since the \ion{Ca}{ii} resonance lines are strongly
  saturated in the line centers, the unsatisfactory description of the
  line core does not affect our results.
\end{itemize}

\begin{figure}
  \includegraphics[width=0.45\textwidth]{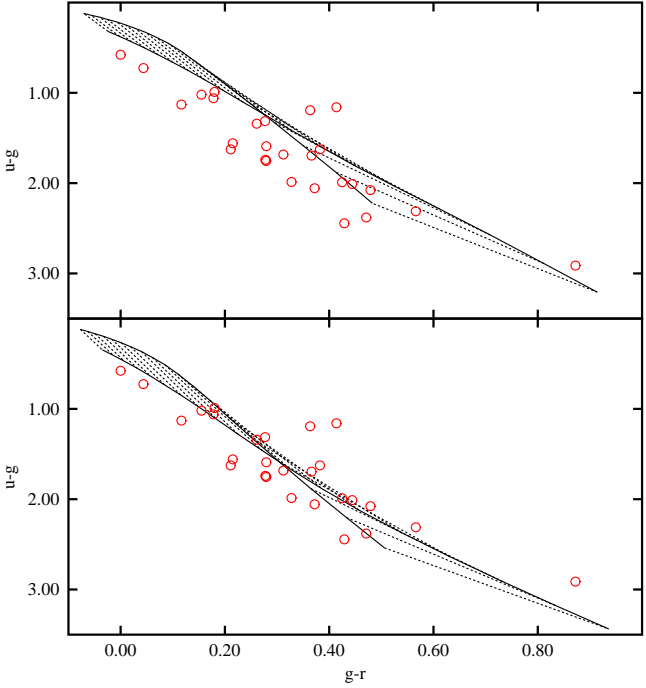} 
\caption{Observed and model colors in the $u-g$ vs. $g-r$ two-color
  diagram. Top panel: the continuous lines are model values for a
  temperature sequence from 5400~K (lower right) to 9000~K (upper
  left) with steps of 200~K. The dotted lines connect the same
  \Teff\ for 2 different abundances; the upper line in the right half
  of the figure corresponds to abundances of SDSS0916+2540, the other
  line to SDSS1103+4144. The (red) circles are the observed data.
  Bottom panel: the same models, but with the neutral broadening
  constant for the \ion{Mg}{ii} resonance lines increased by a factor
  of 10 to simulate a stronger blanketing effect in the near
  ultraviolet.
\label{ugr1}}
\end{figure}

\begin{table}
\caption{Lines identified in SDSS0916+2540} 
\label{lineids}      
\centering          
\begin{tabular}{l c}     
\hline\hline       
\noalign{\smallskip}
 Ion     & Wavelengths  \\ 
         &    [\AA]     \\
\hline    
\noalign{\smallskip}
\ion{Ca}{i}  & 4227.918, 5590.301, 5596.015, 5600.034\\
     & 6104.413, 6123.912, 6163.878, 6440.855\\ 
     & 6464.353, 6495.575, 7150.121, 7204.185\\
\ion{Ca}{ii} & 3934.777, 3964.592, 8544.438, 8664.520\\
\ion{Mg}{i}  & 5168.761, 5174.125, 5185.047\\
\ion{Fe}{i}  & 4384.775, 4405.987, 5271.004, 5271.823\\ 
     & 5329.521, 5372.982, 5398.627, 5407.277\\ 
     & 5431.205, 5448.430\\
\ion{Ti}{i}  & 4534.512, 4536.839, 4537.190, 4537.313\\ 
     & 4983.121, 4992.457, 5000.897, 5008,607\\ 
     & 5015.586, 5015.675, 5041.362, 5066.065\\
\ion{Cr}{i}  & 5207.487, 5209.875\\
\ion{Na}{i}  & 5891.583, 5897.558\\
\hline                  
\end{tabular}
\tablefoot{Some lines, in particular of \ion{Ti}{i}, are strongly
  blended and not identified individually. The wavelengths are on the
  vacuum scale.}
\end{table}

\subsection{Atmospheric parameters from photometry}
By iteration through trial and error we have determined \Teff\ and
abundances and calculated a model, which gives a fairly good fit to
the spectrum and photometry of the most feature-rich object,
SDSS0916+2540. The parameters for this object are \Teff\ =
  5500~K, $\log$[Mg/He] = $-6.9$, $\log$[Ca/He] = $-7.6$,
  $\log$[Fe/He]= $-7.1$ (number abundance ratios).  We then calculated
  a sequence of models with these abundances from \Teff\ = 5000~K to
  ~7000~K in steps of 100~K, and continued up to 9000~K in steps of
  200~K, always assuming a surface gravity $\log g = 8.0$. Similarly a
  second sequence was calculated with abundances appropriate for
  SDSS1103+4144 (\Teff\ = 5700~K, $\log$[Mg/He] = $-7.9$,
    $\log$[Ca/He] = $-9.7$, $\log$[Fe/He] = $-8.1$). The choice of this
  second object is somewhat arbitrary and was motivated by being most
  distant from the first sequence in $(u-g)$ vs. $(g-r)$ color space.

Convolving the spectral energy distribution with the $ugriz$
band-passes of the SDSS photometric system, we constructed the
two-color $(u-g)$ vs. $(g-r)$ diagram shown in Fig.~\ref{ugr1}. Two
conclusions can be drawn from this diagram
\begin{itemize}
\item The location of the theoretical models depends in a complicated
  way on metal abundance, which is caused by an interplay of direct
  absorption in the optical range, and an indirect effect of the
  blanketing by very strong absorption in the ultraviolet. It would be
  difficult to estimate abundances from the photometry.
\item The observed objects form a temperature sequence, with some
  scatter caused by different relative abundances of Mg, Ca, and
  Fe. The general shape agrees with the theoretical grid, but the
  observations are shifted downward and to the left. This is a strong
  indication that we still underestimate the blanketing effect of the
  near ultraviolet spectral lines, although all expected strong lines of the
  detected metals, (most importantly Fe, Mg and Ca) are included in our
  models. 
\end{itemize}

\begin{table*}
\caption{Effective temperature and element abundances for the sample \label{tabresults}}
\centering          
\begin{tabular}{l r r r r r r r}     
\hline\hline    
\noalign{\smallskip}   
 SDSS &\Teff(phot) &\Teff & Mg & Ca &  Fe & Na & other\\ 
       &  [K]       & [K]  &                           \\
\hline                    
\noalign{\smallskip}
0143+0113& 6600 & 6600 & $-$7.4 & $-$9.2 & $-$7.7 &\\
0157+0033& 6200 & 6300 & $-$7.1 & $-$8.1 & $-$7.5 &\\
0205+2155& 6000 & 6000 &      &      &      &      & DQ 6000K C $-$7.0\\
0916+2540& 5500 & 5500 & $-$6.9 & $-$7.6 & $-$7.1 & $-$9.3 & Ti $-$9.0, Cr $-$8.9\\
0925+3130& 5900 & 5900 & $-$7.9 & $-$9.3 & $-$8.0 & $-$9.3&\\
0937+5228& 7000 & 6900 & $-$6.8 & $-$8.3 & $-$7.5 &     &\\
0956+5912& 8600 & 8800 & $-$5.5 & $-$7.3 & $-$6.5 & $-$7.0& H $-$3.25 \\
1038$-$0036&8300& 8200 & $-$6.6 & $-$7.7 & $-$7.2 & $-$8.5 &H $<$ $-$4.7\\
1040+2407& 6100 & 6200 & $-$7.1 & $-$8.0 & $-$7.2 & $-$8.5 &\\
1043+3516& 5800 & 5900 & $-$8.1 &$-$10.1 & $-$7.8 &      &\\
1103+4144& 5800 & 5700 & $-$7.9 & $-$9.7 & $-$8.1 & $-$9.0 &\\
1144+1218& 5400 & 5300 & $-$8.2 & $-$9.9 & $-$8.6 &      &\\
1152+1605& 6900 & 6600 & $-$7.3 & $-$8.5 & $-$7.4 &      &\\
1234+5208& 7800 & 7800 & $-$5.9 & $-$7.5 & $-$6.5 & $-$8.2 & Cr $-$8.2\\
1330+3029& 6200 & 6440 & $-$6.9 & $-$8.1 & $-$7.0 & $-$8.4 &H $\approx$-3\\
1336+3547& 6600 & 6700 & $-$7.1 & $-$8.8 & $-$7.6 & $-$8.7 & H $<$ $-$4.2\\
1356+2416& 6100 & 6100 & $-$7.7 & $-$9.3 &      &      &\\
1404+3620& 6600 & 5900 & $-$7.9 & $-$9.6 & $-$8.6 &      &\\
1405+1549& 7700 & 7700 & $-$6.9 & $-$8.1 & $-$7.0 &      &\\ 
1421+1843& 7200 & 6800 & $-$6.9 & $-$8.2 & $-$7.1 & $-$8.3 \\
1430$-$0151&5700& 6300 & $-$6.4 & $-$7.6 & $-$6.8 & $-$7.9 & Ti $-$8.8, Cr $-$8.4\\
1524+4049& 6000 & 5700 & $-$7.9 & $-$9.6 & $-$7.9 & $-$9.4 \\
1535+1247& 5600 & 6000 & $-$7.3 & $-$8.8 & $-$7.6 & $-$8.6 & Ti $-$9.50, Cr $-$9.10\\
1546+3009& 7200 & 6700 & $-$7.2 & $-$8.5 & $-$7.0 &      \\
1554+1734& 7000 & 7000 & $-$7.4 & $-$8.7 & $-$7.7 & $-$8.5 \\
1616+3303& 6900 & 6700 & $-$6.8 & $-$8.5 & $-$7.0 &      \\
2340+0817& 6300 & 6300 & $-$7.5 & $-$8.4 & $-$7.5 &      \\
\hline                  
\end{tabular}
\tablefoot{Column 2 is the temperature obtained from photometry as
  explained in the text, column 3 is the final adopted
  temperature. The abundances are logarithmic ratios to helium by
  number.}
\end{table*}

\begin{table*}
\caption{Comparison of observed SDSS photometry with theoretical
  colors for the final model \label{compcol}}
\centering          
\begin{tabular}{l r r r r r | l r r r r r} 
\hline\hline    
\noalign{\smallskip}   
 SDSS     &  $g$   & $u-g$ & $g-r$ & $r-i$ & $r-z$& SDSS     &  $g$   & $u-g$ & $g-r$ & $r-i$ & $i-z$\\
\hline                    
\noalign{\smallskip}
0143+0113& 19.380  & 0.990 & 0.180 &$-$0.060 &$-$0.150& 1330+3029& 16.310 & 1.950 & 0.425 &$-$0.012 &$-$0.218\\
          &        & 0.886 & 0.207 &$-$0.061 &$-$0.123&          &        & 1.493 & 0.362 &$-$0.041 &$-$0.151\\
          &        & 0.061 & 0.014 & 0.022 & 0.063&          &        & 0.027 & 0.020 & 0.027 & 0.035\\
0157+0033& 19.566  & 1.655 & 0.366 &$-$0.037 &$-$0.135& 1336+3547& 17.854 & 1.586 & 0.211 &$-$0.074 &$-$0.129\\
          &        & 1.263 & 0.442 &$-$0.059 &$-$0.164&          &        & 0.865 & 0.225 &$-$0.068 &$-$0.136\\
          &        & 0.098 & 0.026 & 0.032 & 0.059&          &        & 0.031 & 0.020 & 0.030 & 0.032\\
0916+2540& 18.346  & 2.873 & 0.872 & 0.089 &$-$0.164& 1356+2416& 18.998 & 1.313 & 0.277 &$-$0.021 &$-$0.092\\
          &        & 2.727 & 0.932 & 0.093 &$-$0.193&          &        & 0.949 & 0.329 &$-$0.031 &$-$0.103\\
          &        & 0.103 & 0.020 & 0.023 & 0.025&          &        & 0.063 & 0.030 & 0.025 & 0.048\\
0925+3130& 18.983  & 2.017 & 0.372 &$-$0.021 &$-$0.136& 1404+3620& 18.818 & 1.152 & 0.364 &$-$0.065 &$-$0.192\\
          &        & 1.471 & 0.383 &$-$0.011 &$-$0.095&          &        & 1.103 & 0.358 &$-$0.012 &$-$0.088\\
          &        & 0.107 & 0.024 & 0.021 & 0.047&          &        & 0.050 & 0.034 & 0.028 & 0.039\\
0937+5228& 19.439  & 1.302 & 0.261 &$-$0.123 &$-$0.089& 1405+1549& 18.940 & 1.020 & 0.155 &$-$0.115 &$-$0.224\\
          &        & 0.754 & 0.233 &$-$0.086 &$-$0.161&          &        & 0.561 & 0.111 &$-$0.112 &$-$0.189\\
          &        & 0.073 & 0.035 & 0.029 & 0.059&          &        & 0.047 & 0.027 & 0.026 & 0.045\\
0956+5912& 18.394  & 0.538 &$-$0.005 &$-$0.155 &$-$0.209& 1421+1843& 20.355 & 1.701 & 0.278 &$-$0.116 &$-$0.156\\
          &        & 0.332 &$-$0.003 &$-$0.134 &$-$0.219&          &        & 1.011 & 0.270 &$-$0.071 &$-$0.163\\
          &        & 0.030 & 0.027 & 0.029 & 0.044&          &        & 0.171 & 0.039 & 0.044 & 0.116\\
1038$-$0036& 16.984& 0.686 & 0.044 &$-$0.146 &$-$0.242& 1430$$-$$0151&19.475& 1.970 & 0.444 & 0.038 &$-$0.117\\
          &        & 0.459 & 0.100 &$-$0.147 &$-$0.235&          &        & 1.681 & 0.540 &$-$0.021 &$-$0.181\\
          &        & 0.029 & 0.025 & 0.019 & 0.021&          &        & 0.143 & 0.030 & 0.028 & 0.064\\
1040+2407& 19.337  & 2.341 & 0.471 &$-$0.005 &$-$0.251& 1524+4049& 20.135 & 1.586 & 0.382 &$-$0.013 &$-$0.063\\
          &        & 1.515 & 0.470 &$-$0.036 &$-$0.162&          &        & 1.846 & 0.430 & 0.004 &$-$0.081\\
          &        & 0.149 & 0.030 & 0.029 & 0.074&          &        & 0.123 & 0.035 & 0.034 & 0.083\\
1043+3516& 18.959  & 1.643 & 0.312 & 0.036 &$-$0.074& 1535+1247& 15.977 & 2.037 & 0.479 & 0.045 &$-$0.072\\
          &        & 1.769 & 0.335 &$-$0.015 &$-$0.085&          &        & 1.573 & 0.425 &$-$0.015 &$-$0.113\\
          &        & 0.079 & 0.029 & 0.026 & 0.041&          &        & 0.031 & 0.027 & 0.028 & 0.028\\
1103+4144& 19.731  & 1.947 & 0.328 & 0.019 &$-$0.060& 1546+3009& 19.807 & 1.551 & 0.279 &$-$0.119 &$-$0.167\\
          &        & 1.677 & 0.392 & 0.021 &$-$0.078&          &        & 1.153 & 0.275 &$-$0.074 &$-$0.147\\
          &        & 0.188 & 0.034 & 0.029 & 0.061&          &        & 0.088 & 0.027 & 0.030 & 0.067\\
1144+1218& 18.412  & 2.271 & 0.566 & 0.091 &$-$0.024& 1554+1734& 17.599 & 1.060 & 0.178 &$-$0.048 &$-$0.202\\
          &        & 2.014 & 0.559 & 0.053 & 0.041&          &        & 0.648 & 0.171 &$-$0.073 &$-$0.149\\
          &        & 0.087 & 0.026 & 0.027 & 0.029&          &        & 0.028 & 0.018 & 0.016 & 0.026\\
1152+1605& 20.169  & 1.519 & 0.215 &$-$0.058 &$-$0.111& 1616+3303& 19.252 & 1.714 & 0.279 &$-$0.096 &$-$0.104\\
          &        & 1.042 & 0.289 &$-$0.068 &$-$0.146&          &        & 1.120 & 0.286 &$-$0.075 &$-$0.145\\
          &        & 0.133 & 0.032 & 0.037 & 0.097&          &        & 0.096 & 0.026 & 0.027 & 0.049\\
1234+5208& 18.415  & 1.090 & 0.117 &$-$0.125 &$-$0.278& 2340+0817& 20.248 & 2.445 & 0.429 &$-$0.052 &$-$0.102\\
          &        & 0.755 & 0.179 &$-$0.122 &$-$0.206&          &        & 1.301 & 0.392 &$-$0.055 &$-$0.147\\
          &        & 0.049 & 0.030 & 0.023 & 0.035&          &        & 0.274 & 0.028 & 0.034 & 0.080\\

\hline                
\end{tabular}
\tablefoot{The observed values were corrected by $-$0.04 ($u$)
  and 0.02 ($z$) to transform them to the AB magnitude scale (adopting
  the corrections given on the SDSS DR7 web pages). For each object
  the first line are the observed colors, the second the model fit,
  and the third the errors of the observations.}
\end{table*}

\begin{figure*}
  \includegraphics[width=0.85\textwidth]{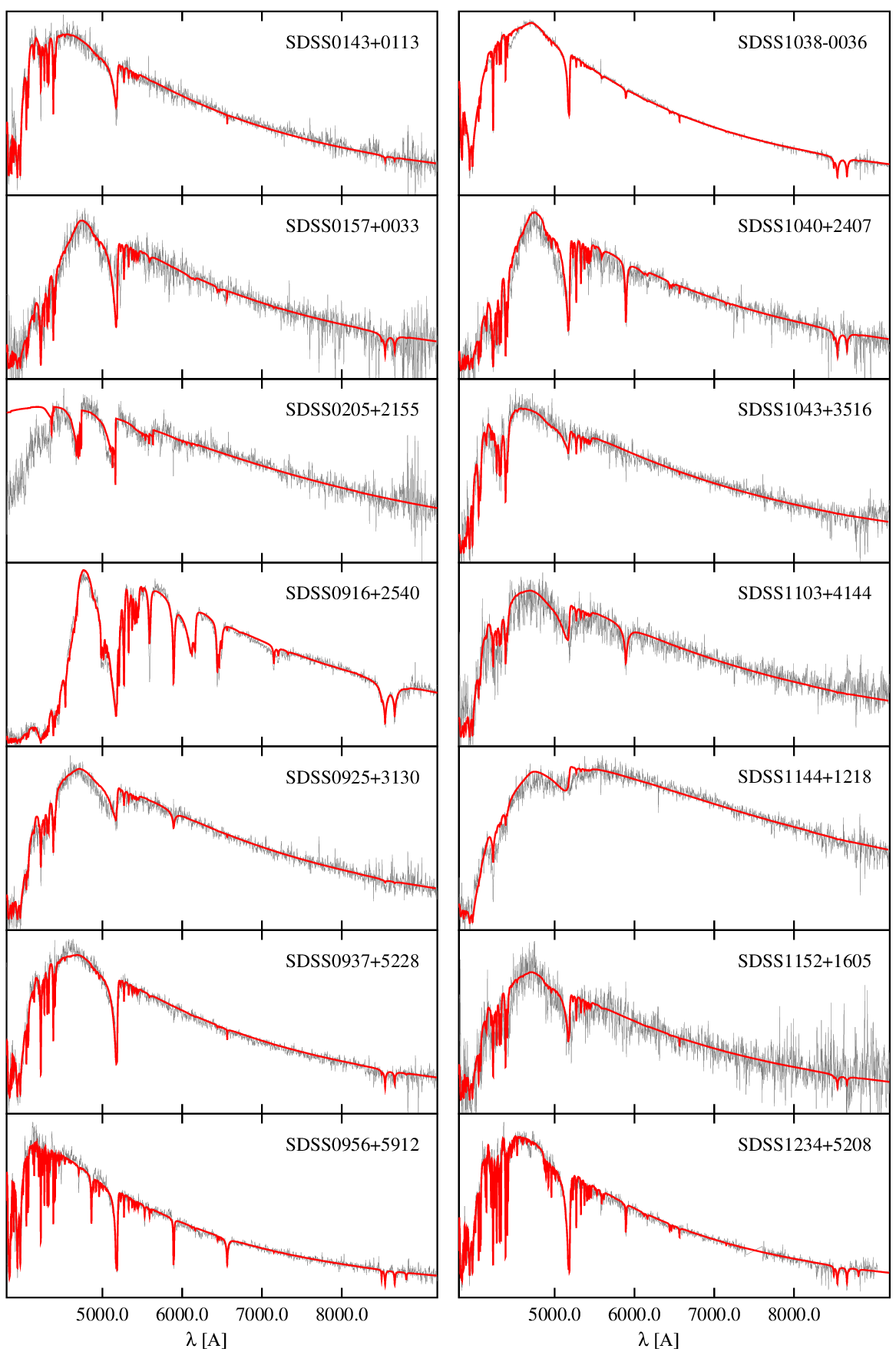} 
\caption{Observed spectra (light grey) and theoretical models
  (dark/red lines). Both spectra were convolved with a Gaussian with
  3\AA\ FWHM corresponding to the resolution of the SDSS spectra. The
  vertical axis is the flux $F_\lambda$ on a linear scale, with zero
  at the bottom.
\label{fitplota}}
\end{figure*}

\begin{figure*}
  \includegraphics[width=0.86\textwidth]{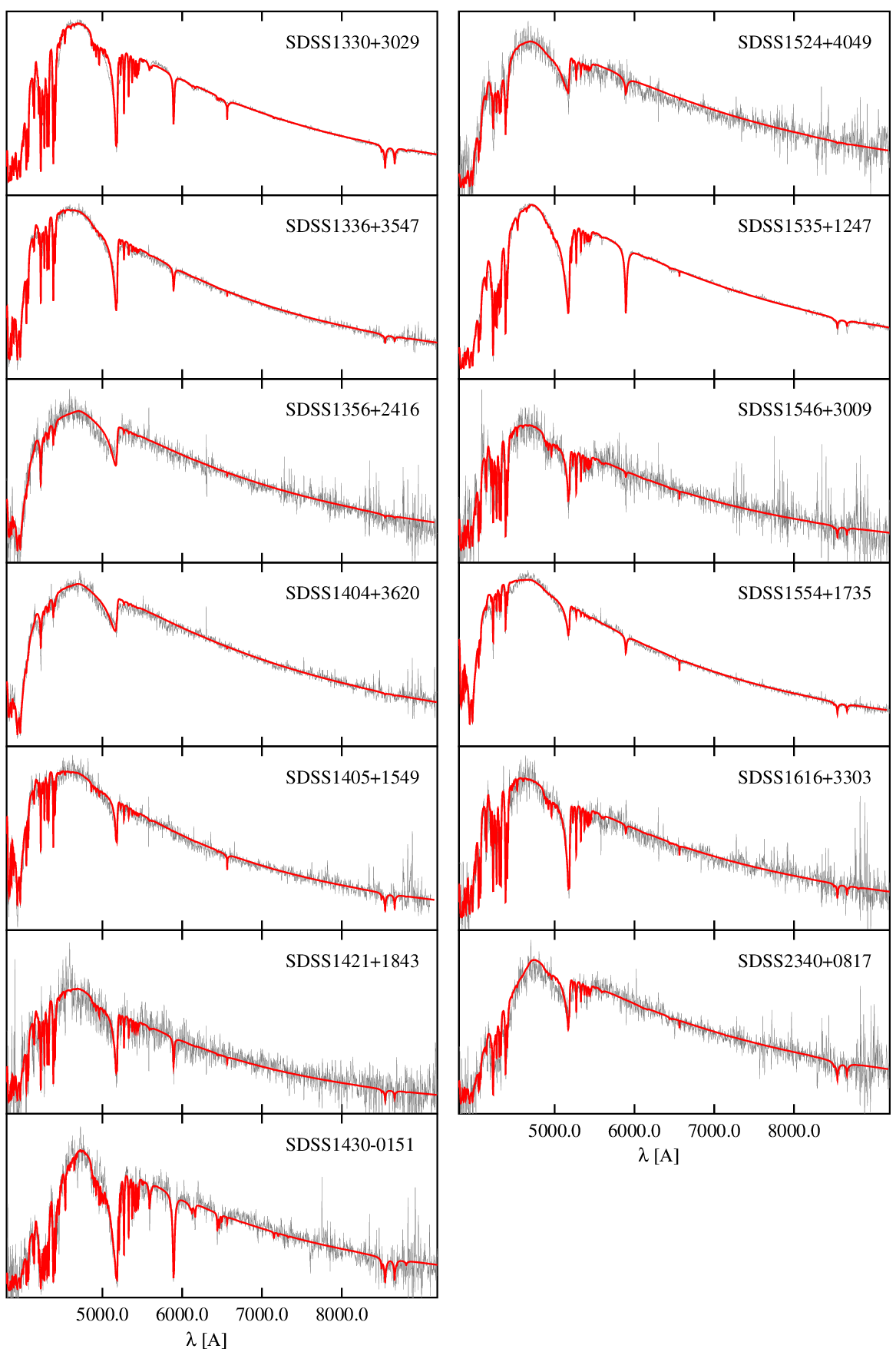} 
\caption{The same as Fig.~\ref{fitplota} for the second half of the sample.
\label{fitplotb}}
\end{figure*}

The \ion{Mg}{i}/\ion{Mg}{ii} resonance lines are predicted to be
extremely strong, and it is well known that the simple impact
approximation fails to reproduce the spectra in detail
\citep{Zeidler-KT.Weidemann.ea86}. Potential energy curves for the
MgII-He system were calculated by \cite{Monteiro.Cooper.ea86},
but are not available in tabular form in the literature. Since no
ultraviolet spectra exist for our objects, a comparison with
observations is not currently possible.

In order to study the effect of stronger blanketing, we have increased
the neutral broadening constant $\Gamma_6$ for the \ion{Mg}{ii}
resonance lines by a factor of 10, which was shown by
\cite{Zeidler-KT.Weidemann.ea86} to produce a line consistent with the
Mg abundance derived from the optical spectrum.  The resulting
two-color diagram is shown in the lower panel in Fig.~\ref{ugr1}. The
new sequence is a better fit to the observations, although the
blanketing is still underestimated at the hot end of the sequences.

Since metals not observable in the optical spectra might contribute to
the free electrons as well as to the absorption, we made a few
test calculations. We added the elements C, O, Al, Si, P, S, Sc, V,
Mn, Co, Ni, Cu, and Zn with abundance ratios relative to Mg taken from
\cite{Klein.Jura.ea10}, if observed in GD40, or with solar ratios
otherwise. In another test series we used only the observed metals,
but added 372 bound-free cross sections, mostly from \ion{Fe}{i}.
These test models showed no significant differences in the optical
spectra and we therefore used only the identified elements in the
analysis of the sample. The only exception is hydrogen, for which
$\log$[H/He] = $-4$ (number abundances) was used unless otherwise
noted in Table~\ref{tabresults}. This is a fairly typical value in the
cooler objects of the \cite{Dufour.Bergeron.ea07} sample. It is also
the upper limit in our spectra with higher signal-to-noise ratios, and
a further reduction does not change the models anymore, since the
electrons come predominantly from the metals.

As the next step we used the theoretical $ugriz$ magnitudes for
both sequences (with nominal broadening constants) to estimate an
effective temperature for all objects in the sample using a $\chi^2$
minimization. The second parameter in the fitting procedure was a
number set to 1 for one sequence and 0 for the other. This allowed for
an approximate interpolation between the two sets of abundances.
The resulting temperature is given in Table~\ref{tabresults} as
\Teff (phot) in the second column. Assuming that \logg\ = 8, this
fitting procedure also gives a photometric distance. More than half
of our sample are closer than 100~pc, and all closer than
200~pc. Interstellar reddening should not be important, compared to
the other sources of uncertainty, and was neglected in our fits.

\subsection{Spectral fitting}
We used the temperature of the grid point closest to the best-fit
photometric temperature as a starting value in the analysis of the
individual objects.  The abundances were varied, following a visual
comparison of observed and theoretical spectrum. A small grid with
\Teff\ 100-300~K above and below the current best value was calculated
and theoretical SDSS colors for these models compared to the observed
ones. Most weight was given to the most accurate colors, i.e. $g-r$
and $r-i$. This procedure was iterated, until a reasonable fit was
achieved to the spectrum and the colors. $g-r$, $r-i$, and $i-z$ were
usually reproduced by the final model within 1-2 $\sigma$; for $u-g$
the error was often (not always) larger (see Table~\ref{compcol}),
because of missing absorption in the $u$ region. In the higher quality
spectra we could additionally use the ionization equilibrium of
calcium to determine the temperature. In a few cases the final
temperature differs by several 100~K from the photometric starting
value, because two objects with similar colors in the intermediate
range of Fig.~\ref{ugr1} can have very different temperatures,
depending on the details of the metal contamination. Figures for all
spectra with the fits are displayed in Fig.\ref{fitplota} and
Fig.\ref{fitplotb}.

\begin{figure}
  \includegraphics[width=0.48\textwidth]{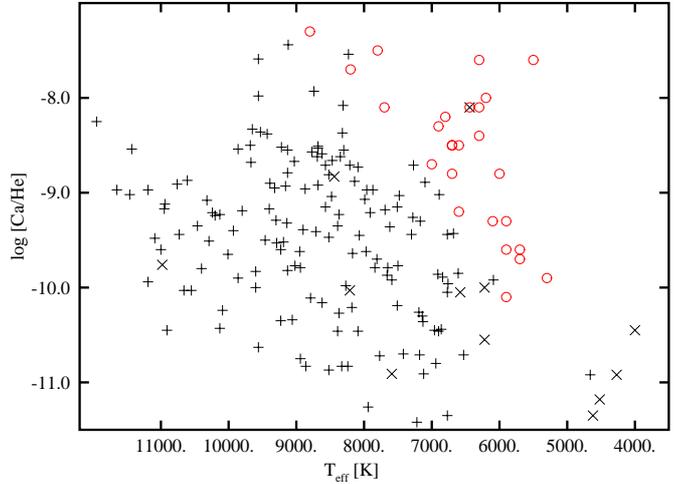} 
\caption{Calcium abundances from \cite{Dufour.Bergeron.ea07} as
  displayed in their Fig.~9, with our present results added. Crosses
  are for objects found in the SDSS, x for objects from the
  \cite{Bergeron.Leggett.ea01} sample, with parameters redetermined by
  \cite{Dufour.Bergeron.ea07}. Our new results are the (red) open
  circles.
\label{Dufourfig}}
\end{figure}

\section{Results}
The results for \Teff\ and the abundances of the observed elements are
given in Table~\ref{tabresults}. The objects of our sample form a
natural extension of the DZ temperature sequence studied by
\cite{Dufour.Bergeron.ea07}. We reproduce their results of the Ca/He
abundances as a function of \Teff\ in Fig.~\ref{Dufourfig}, with our
results added as red open circles. The new DZ stars discovered here
nicely fill in the gap which was apparent in the the
\citeauthor{Dufour.Bergeron.ea07} sample between 5500 and 8500~K at
higher abundances, $\log$[Ca/He] = $-$8 to $-$10. In fact, our new sample
of DZ white dwarfs provides several more objects at very large Ca/He
around 6000~K, occupied up to now by the lonely G165-7. The fact that
our sample does not include objects with low Ca/He abundances is
likely a result of our selection procedure.

\subsection{Error estimates}
The $\chi^2$ fitting of the photometry described above gives formal
errors for the effective temperatures in the range from 50 to
800~K. These are, however, not meaningful at all, as the metal
abundances vary considerably from object to object, and differ
generally from those used for the two sequences of our model grid. The
iterative procedure used to determine individual parameters does not
lead to formal error estimates. From the visual inspection we find
that the fit gets noticeable worse for a change of 150~K, if the
spectrum has a good signal-to-noise ratio (S/N), as well as many clear
spectral features, in particular lines of \ion{Ca}{i} and
\ion{Ca}{ii}. We hence adopt 150~K as our error estimate for the best
cases, and estimate an uncertainty of 300~K for the lowest S/N objects
(where the surface gravity is always fixed at the canonical value of
\logg\ = 8).

If the temperature and gravity are fixed and their errors neglected,
the uncertainties of the abundances are estimated at 0.1-0.2 dex for
Mg, Ca, and Na, and 0.2-0.3 dex for Fe, Ti, Cr. If possible errors of
\Teff\ and variations of surface gravity are included, the abundance
errors can probably be up to twice these values in the worst case.

The comparison of observed and calculated $u-g$ colors shows that in
most cases our models are still missing absorption near 3500~\AA. In
our simulation with enhanced \ion{Mg}{ii} broadening constant we found
that the model flux is enhanced over the whole observed optical
range. However, the slope of the continuum and the spectral features
do not change significantly. We are therefore confident that this
problem does not add a larger uncertainty to the derived abundances.

\begin{table*}
\caption{Depth of the convection zone and diffusion time scales \label{difftimes}}
\centering          
\begin{tabular}{r r r r r r r r r }     
\hline\hline    
\noalign{\smallskip}   
 SDSS       &   q  & \Teff  &   Na  &  Mg   &  Ca   &  Ti   &   Cr  & Fe \\ 
             &      &  [K]   &   \\
\hline                    
\noalign{\smallskip}
  0143+0113 & $-$5.3184 & 6600 & 5.908 & 5.916 & 5.840 & 5.802 & 5.824 & 5.805\\
  0157+0033 & $-$5.4374 & 6300 & 5.820 & 5.833 & 5.736 & 5.713 & 5.718 & 5.698\\
  0916+2540 & $-$5.6914 & 5500 & 5.633 & 5.654 & 5.530 & 5.517 & 5.496 & 5.472\\
  0925+3130 & $-$5.4124 & 5900 & 5.879 & 5.896 & 5.799 & 5.790 & 5.789 & 5.769\\
  0937+5228 & $-$5.3260 & 6900 & 5.903 & 5.906 & 5.828 & 5.781 & 5.798 & 5.783\\
  0956+5912 & $-$5.2882 & 8800 & 5.972 & 5.957 & 5.872 & 5.808 & 5.802 & 5.810\\
  1038$-$0036&$-$5.1065 & 8200 & 6.069 & 6.080 & 6.014 & 5.957 & 5.938 & 5.952\\ 
  1040+2407 & $-$5.4839 & 6200 & 5.774 & 5.788 & 5.684 & 5.668 & 5.667 & 5.646\\
  1043+3516 & $-$5.4041 & 5900 & 5.885 & 5.902 & 5.806 & 5.797 & 5.797 & 5.777\\
  1103+4144 & $-$5.4381 & 5700 & 5.870 & 5.890 & 5.789 & 5.788 & 5.779 & 5.757\\
  1144+1218 & $-$5.4411 & 5300 & 5.918 & 5.943 & 5.841 & 5.852 & 5.835 & 5.810\\
  1152+1605 & $-$5.3484 & 6600 & 5.894 & 5.902 & 5.820 & 5.782 & 5.803 & 5.782\\
  1234+5208 & $-$5.2821 & 7800 & 5.961 & 5.946 & 5.867 & 5.808 & 5.811 & 5.812\\
  1330+3029 & $-$5.4610 & 6440 & 5.799 & 5.810 & 5.706 & 5.679 & 5.683 & 5.662\\
  1336+3547 & $-$5.3139 & 6700 & 5.924 & 5.930 & 5.853 & 5.811 & 5.831 & 5.813\\
  1356+2416 & $-$5.3667 & 6100 & 5.903 & 5.918 & 5.830 & 5.811 & 5.819 & 5.800\\
  1404+3620 & $-$5.3800 & 5900 & 5.907 & 5.923 & 5.832 & 5.822 & 5.825 & 5.805\\
  1405+1549 & $-$5.1901 & 7700 & 6.026 & 6.016 & 5.950 & 5.893 & 5.888 & 5.890\\
  1421+1843 & $-$5.3542 & 6800 & 5.879 & 5.885 & 5.802 & 5.760 & 5.779 & 5.760\\
  1430$-$0151&$-$5.5399 & 6300 & 5.734 & 5.748 & 5.634 & 5.616 & 5.611 & 5.589\\
  1524+4049 & $-$5.4519 & 5700 & 5.859 & 5.878 & 5.776 & 5.774 & 5.764 & 5.742\\
  1535+1247 & $-$5.4688 & 6000 & 5.803 & 5.819 & 5.717 & 5.707 & 5.703 & 5.682\\
  1546+3009 & $-$5.3588 & 6700 & 5.876 & 5.884 & 5.801 & 5.761 & 5.781 & 5.761\\
  1554+1734 & $-$5.2442 & 7000 & 5.978 & 5.973 & 5.908 & 5.856 & 5.870 & 5.864\\
  1616+3303 & $-$5.3819 & 6700 & 5.862 & 5.870 & 5.782 & 5.743 & 5.762 & 5.740\\
  2340+0817 & $-$5.4007 & 6300 & 5.858 & 5.870 & 5.779 & 5.754 & 5.762 & 5.742\\ 
\hline                  
\end{tabular}
\tablefoot{$q$ is the fractional depth of the convection zone, $q =
  \log M_{cvz}/M$. Columns 4 to 9 give the logarithm of the diffusion
  time scale $\tau$ for the elements Na to Fe in years.}
\end{table*}

\subsection{Notes for individual objects}
\noindent
{\bf SDSS0157+0033:} The \ion{Mg}{i} triplet lines seem to be split and
shifted, suggesting a noticeable magnetic field for this white
dwarf. Higher S/N data are needed to confirm this hypothesis.

\noindent
{\bf SDSS0205+2155:} This is a DQ with Swan bands of the C$_2$
molecule. From comparison with a grid of DQ models we estimate an
effective temperature of about 6000~K, and $\log$ [C/He] = $-$7.0.  The
decrease of the flux at the blue end may be due to calcium and
possibly iron, which is an interesting possibility, since the current
explanation for the DQ stars is dredge-up of carbon from deeper layers
\citep{Koester.Weidemann.ea82, Pelletier.Fontaine.ea86}. We estimate
that the Ca abundance would need to be around $\log$[Ca/He] = $-10.0$ to
$-10.5$, but this is pure speculation at the moment.

\noindent
{\bf SDSS0937+5228:} This is a previously known white dwarf
(WD0933+526), first identified as a DZ by \cite{Harris.Liebert.ea03}.

\noindent
{\bf SDSS0956+5912:} \cite{Dufour.Bergeron.ea07} find \Teff\ = 8230~K
for this star, whereas we find a somewhat higher temperature from the
SDSS photometry, continuum slope as well as blue flux. Our Ca
abundance is also higher by about a factor two.

\noindent
{\bf SDSS1038$-$0036:} Our temperature is significantly higher than the
6770~K found by \cite{Dufour.Bergeron.ea07}. As can be seen in their
Fig.~7, the \ion{Ca}{ii} lines near 8600~\AA\ are much too weak in the
model. It seems that a higher temperature is a better fit. The
spectrum from DR7 shows unexplained features near 4720 and 4865~\AA;
we have used the reduction of DR4 for the same spectrum, where the
features are absent.

\noindent
{\bf SDSS1144+1218:} The model fits the photometry very well, but the
slope and absolute scale of the observed spectrum disagree. We
consider this to be a calibration problem of the spectrum and the
model in Fig~\ref{fitplotb} is adjusted with a linear correction.

\noindent
{\bf SDSS1152+1605:} Although the spectrum is very noisy there are
some indications of possible Zeeman splitting in the stronger lines.

\noindent
{\bf SDSS1330+3029:} This star is also known as WD\,1328+307 and
G\,165-7. It was analyzed by \cite{Dufour.Bergeron.ea06} who found it
to be a weakly magnetic ($\approx 650$kG) DZ white dwarf with Zeeman
splitting in lines of Ca, Na, and Fe. We used their parameters
\Teff\ and \logg, as well as their abundances and calculated a model
           {\em without considering a magnetic field}. As expected,
           the lines are slightly weaker and narrower in our model,
           due to the absence of the splitting, but otherwise it
           agrees very well with their results.

\noindent
{\bf SDSS1404+3620:} The photometric temperature is very likely too
high. The continuum slope and \ion{Ca}{i}/\ion{Ca}{ii} ionization
demand a much lower \Teff, which we have adopted here. The theoretical
colors from the final model agree with the observations.

\noindent
{\bf SDSS1535+1247:} This is a known white dwarf (WD\,1532+129,
G137-24), which was classified as DZ by
\cite{Kawka.Vennes.ea04}. Because the spectrum is similar to that of
G165-7, they adopted the model of \cite{Wehrse.Liebert80} for that
star with \Teff\ = 7500~K and found metal abundances of $\approx
1/100$ solar. They note, however, that a fit to the $VJH$ photometry
results in \Teff\ = 6000$\pm$400~K.  The $u-g$ and $g-r$ colors are
similar to SDSS1330+3029, yet the slope of the spectra is quite
different. Also, adding the $i$ and $z$ magnitude for the photospheric
fit results in significantly lower temperature. The best compromise
using the resonance and excited lines from the two ionization stages
is \Teff\ = 6000~K, in agreement with the photometric result of
\cite{Kawka.Vennes.ea04}. The theoretical $griz$ photometry for the
final model agrees well with the observations (Table~\ref{compcol}).

\noindent
{\bf SDSS1546+3009:} This object may be weakly magnetic, with
splittings and shifts apparent in many metal lines. A higher S/N
spectrum is needed for confirmation.

\section{Element abundances, diffusion, and accretion}
Heavy elements in a helium-dominated atmosphere will sink out of the
outer, homogeneously mixed convection zone into deeper layers. The
abundance observed depends on the interplay of accretion from the
outside and diffusion at the bottom of the convection zone. These
diffusion time scales can be calculated for all objects using the
methods and input physics as described in \cite{Koester.Wilken06} and
\cite{Koester09}. The data are collected in Table~\ref{difftimes}.

The size of the convection zone, and the diffusion time scales depend
on the effective temperature, which determines the stellar
structure. In addition it depends on the metal composition of the
atmosphere, because the atmospheric data at Rosseland optical depth 50
are used as outer boundary conditions for the envelope calculations.
However, the total range of time scales over all objects and all
elements only varies within a factor of $\approx 4$, from
$3\,10^5$ to $1.2\,10^6$ years.

As discussed in \cite{Koester09}, the interpretation of observed
abundances, and their relation to the abundances in the accreted
material depends on the identification of the current phase within the
accretion/diffusion scenario: initial accretion, steady state, or
final decline. Except for the case of hotter DAZ, with diffusion time
scales of a few years or less, we generally do not know in which phase
we observe the star. The currently favored source for the accreted
matter is a dusty debris disk, formed by the tidal disruption of
planetary rocky material. The lifetime of such a debris disk is highly
uncertain; estimates put it around $1.5\,10^5$ yrs \citep{Jura08,
  Kilic.Farihi.ea08}. If the lifetime is really that short, the steady
state phase would never be reached for the cool DZ analyzed here. The
observable abundances would be close to the accreted abundances during
the initial accretion phase. If the accretion rate declines
exponentially, this abundance pattern could persist for a longer
period. If the accretion is switched off abruptly, the element
abundances would diverge, according to their diffusion time scales.

The differences between the time scales of the four elements Mg, Na,
Ca, Fe, which are observed in most objects, are at most a factor of
1.4. Given the relatively large differences between the diffusion time
scales of Mg and Fe, is possible to attribute the scatter of the Fe/Mg
ratio to differences in the time since the accretion episode?  Let us
assume for a moment that all objects have reached similar abundances
when the accretion stops.  The observed range in Mg abundances of 2.7
dex would, under this assumption, be due to an exponential decline for
a duration of 6.22 $\tau$(Mg). During this time the Mg/Fe ratio would
change by a factor of 12 or 1.1 dex, and we would expect a correlation
between Mg abundance and the Fe/Mg ratio. This model is obviously a
strong oversimplification, but at least in a statistical sense it
might be true that objects with lower overall metal abundance might
have spent more time since the accretion stopped. Is such an effect
observable?

\begin{figure}
  \includegraphics[width=0.45\textwidth]{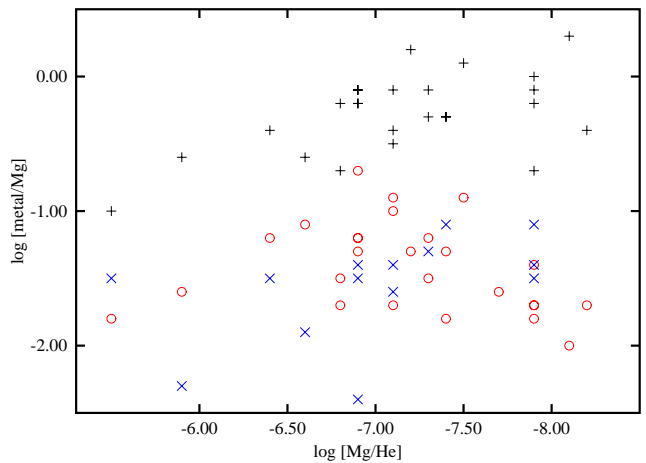} 
\caption{Ca/Mg (red circles), Fe/Mg (black crosses), and Na/Mg (blue
  x) as a function of the magnesium abundance.
\label{ratios}}
\end{figure}

Fig.~\ref{ratios} shows the comparison for the Na/Mg, Ca/Mg, and Fe/Mg
ratios. There is no evolution visible in the first two ratios. For
Na/Mg this is expected, since the diffusion time scales are very
similar and the ratio should always stay close to the accretion
ratio. This would imply that the Na/Mg ratio in the accreted material
can vary by much more than a factor of ten.

If anything, the Fe/Mg ratio shows an increase, in spite of the
shorter time scales of iron. This means that the Mg abundance cannot
be interpreted as an indicator of time since accretion; more likely
the accretion reaches different values in different objects. The
variation by a factor of ten in the Fe/Mg ratio could then be explained
by subsequent diffusion.

The above considerations are rather speculative and need to be taken
with a grain of salt. Abundance errors in individual objects could add
up to 0.5 dex, and the theoretical diffusion time scales use several
approximations \citep{Koester09} -- we cannot exclude that the
uncertainties are as high as a factor of 1.5. Because the time scales
for the different elements given in Table~\ref{difftimes} are very
similar, such small changes could even invert the trends discussed
above.

\begin{table}
\caption{Average logarithmic abundance ratios for Fe/Mg, Ca/Mg,
  Na/Mg, and Na/Ca \label{meandata}}
\centering          
\begin{tabular}{r r r r r}
\hline\hline    
\noalign{\smallskip}   
  Data                 & [Fe/Mg] & [Ca/Mg] & [Na/Mg] & [Na/Ca]  \\
\hline                    
\noalign{\smallskip}
average                &  $-$0.35  &  $-$1.42  &  $-$1.56  & $-$0.19\\
$\sigma$(mean)         &   0.06  &   0.06  &   0.09  &  0.08\\
$\sigma$(distribution) &   0.38  &   0.33  &   0.37  &  0.58\\
bulk Earth             &  $-$0.13  &  $-$1.21  &  $-$1.77  & $-$0.56\\
solar                  &  $-$0.10  &  $-$1.26  &  $-$1.36  & $-$0.10\\
\hline                  
\end{tabular}
\tablefoot{$\sigma$(mean) is the statistical error of the mean value,
  $\sigma$(distribution) is the width of the distribution.}
\end{table}

An alternative interpretation could be to assume that the metal ratios
are all still close to the accreted, assumed to be similar in all
objects, and that the observed distribution is caused by uncertainties
of the abundances and atmospheric parameters
(e.g. \logg). Table~\ref{meandata} collects relevant data for mean
values and distributions of the metal-to-Mg abundance ratios. The
width of the distributions could be completely explained, assuming
typical errors of 0.15 dex for Mg, and 0.25 dex for Fe, Ca, and Na. In
view of the error discussion above this is not implausible.

\section{Conclusions}
We have identified 26 DZ and one DQ stars across and below the main
sequence in the $u-g$ vs. $g-r$ SDSS two-color diagram, most of which
are new detections. A search routine tailored to the specific spectral
features observed in these DZ was needed to avoid the confusion by
the much more numerous main sequence stars and QSOs in this
region. From our analysis with theoretical model atmospheres, which
uses newly developed line profiles for the strongly broadened Ca and
Mg lines, we determined the white dwarf temperatures and abundances
for the polluting elements Ca, Mg, Fe, Na, and in a few cases Cr and
Ti. The temperatures and abundances show that these objects form a
sequence which continues that of the hotter DZs in the sample of
\cite{Dufour.Bergeron.ea07}. The new objects fill the deficiency of
objects with \Teff $\la$ 8000~K at high Ca/He abundances, which was
apparent in that work. We do not find new low-abundance DZ, which is
most likely a consequence of the design of our search algorithm.

Very little is known for these DZ regarding their current phase in the
accretion/diffusion scenario. However, if the accretion is, or was,
from a circumstellar dust disk, it is unlikely that a steady state
phase occurs, since the diffusion time scales are of the same order as
the expected lifetime of such a disk.  Because the diffusion time
scales of the observed elements are very similar, we might expect that
the metal-to-metal ratios are not too far from those in the accreted
matter, and we compare the average values of these elements in the
bulk Earth \citep{Allegre.Poirier.ea95} and the Sun
\citep{Asplund.Grevesse.ea09} in Table~\ref{meandata}. The average
abundance ratios found for our DZ sample are similar to those of
both bulk Earth and the Sun, with the Na/Ca ratio slightly favoring
the solar abundance ratio.

An important element for a distinction between different sources would
be carbon, which is not detected in our sample. This element is about
a factor of 100 under-abundant in the bulk Earth, compared to the Sun
and the interstellar medium. The DQ star SDSS0205+2155 demonstrates
that $\log$[C/He] = $-$7 produces clearly visible Swan bands. However,
with the metal pollution in the other objects, the pressure and the
transparency in the atmosphere are lower (due to more free electrons),
and the typical upper limits are $\log\mathrm{[C/He]}= -5.5$ to
  $-6.0$, which is too high to distinguish between bulk Earth and
  solar/ISM carbon abundances. 

Another critical element for all possible explanations is hydrogen.
We note that in all objects the abundance of hydrogen relative to the
metals is less than the solar value, even in the one case where it is
detected. This is consistent with accretion of predominantly
volatile-depleted material, since hydrogen accreted from the
interstellar matter or from circumstellar water/ice would stay in the
convection zone and can only accumulate with time.

In the case of the DBZ star GD40, \cite{Klein.Jura.ea10} were
able to deduce very detailed conclusions from a comparison of
photospheric abundances with those of various solar system bodies. The
abundance ratios we obtain here roughly follow that of the bulk
Earth. However, the remaining uncertainties of surface gravity,
abundances, and time since the last accretion event do not allow us to
draw any further conclusions in the present study. High resolution,
high S/N observations and extending the spectral coverage at least to
3200~\AA, or better to 2700~\AA\ (to cover the
\ion{Mg}{ii}/\ion{Mg}{i} resonance lines) are needed to refine the
element abundances and detect the very crucial element silicon, and if
possible carbon and oxygen.

We can, however, estimate a lower limit for the total accreted mass
based on the available data for the convection zone. For
SDSS0956+5912, which has the highest metal pollution, we find a total
mass for the {\em observed} metals Mg, Ca, Fe, Na of
$1.5\,10^{23}$g. Adding oxygen with the same ratio to Mg as in GD40
\citep{Klein.Jura.ea10} this number becomes $4.8\,10^{23}$g, even more
than in the extremely heavily polluted SDSS0738+1835
\citep{Dufour.Kilic.ea10}. The total amount of H in the convection
zone is $1.0\,10^{24}$g. For SDSS1144+1218, which has the lowest
abundances, these numbers are $3.2\,10^{20}$g (observed metals), and
$7.9\,10^{20}$g (including oxygen). These masses span the range of the
most massive asteroids in our own planetary system. These are the
absolute minimum of the accreted masses -- depending on how long ago
the accretion ended, these masses can be at least two orders of
magnitude larger, which will require minor planets much larger than
known in the Solar system. Another open question, recently discussed
by \cite{Farihi.Dufour.ea11} in the context of G77-50, a DAZ of
similar low temperature, is how such a very massive asteroid is
suddenly driven into its host star from an orbit apparently stable
over the past 5 Gyrs.

\acknowledgement{This work has made extensive use of VALD, the Vienna
Atomic Line Database \citep{Piskunov.Kupka.ea95,
Ryabchikova.Piskunov.ea97, Kupka.Piskunov.ea99,
Kupka.Ryabchikova.ea00}}

\end{document}